\documentclass[twocolumn,superscriptaddress,amsmath,amssymb,aps,prb]{revtex4}

\usepackage{graphicx}
\usepackage[space]{grffile}
\usepackage{dcolumn}
\usepackage{bm}
\usepackage{subfigure}
\usepackage{mathrsfs}
\usepackage{booktabs}
\usepackage{notoccite}
\usepackage{float}
\usepackage{placeins}
\usepackage{enumerate}
\usepackage{gensymb}

\begin{document}

\title{Quantum--continuum simulation of underpotential deposition \\ at electrified metal--solution interfaces}

\author{Stephen E. Weitzner}
\author{Ismaila Dabo}
\affiliation{Department of Materials Science and Engineering, Materials Research Institute, and Penn State Institutes of Energy and the Environment, The Pennsylvania State University, University Park, PA 16802, USA \\  Email: weitzner@psu.edu}

\begin{abstract}
The underpotential deposition (UPD) of transition metal ions is a critical step in many electrosynthetic approaches. While UPD has been intensively studied at the atomic level, first-principles calculations in vacuum can strongly underestimate the stability of underpotentially deposited metals. It has been shown recently that the consideration of coadsorbed anions can deliver more reliable descriptions of UPD reactions;  however, the influence of additional key environmental factors such as the electrification of the interface under applied voltage and the activities of the ions in solution have yet to be investigated. In this work, copper UPD on gold is studied under realistic electrochemical conditions using a quantum--continuum model of the electrochemical interface.  We report here on the influence of surface electrification, concentration effects, and anion coadsorption on the stability of the copper UPD layer on the gold (100) surface.
\end{abstract}

\maketitle

\section{\label{sec:introduction}Introduction}

Underpotential deposition (UPD) has played an increasingly important role in the electrochemical preparation of nanomaterials with atomically thin metal film coatings for catalysis, imaging, and sensing applications.\cite{Yu2009, Price2011, Personick2011, Yu2013, Kumar2015, Yan2015}  The UPD process is characterized by the formation of a (sub)monolayer of metal ions on a more noble metal substrate in a voltage range more positive than the reversible reduction potential of the adsorbing ion.  The voltage at which the adlayer desorbs from the surface during an anodic scan is typically referenced to the bulk stripping potential of the adsorbed metal film and is termed the underpotential shift ($\Delta \Phi_\text{upd}$). Kolb and co-workers correlated underpotential shifts with differences in work functions of the substrate and the depositing metal (Fig.~1) suggesting that a charge transfer between the adlayer and the substrate may account for the larger adsorption energy of the adatom on the foreign surface.\cite{Kolb1974} Since their seminal work, numerous studies have been performed in an effort to characterize a wide variety of UPD couples with the aim of understanding the voltammetric dependence of the formation and stability of the UPD layer in addition to its composition and surface structure.\cite{Möller1995, Ikemiya1995, Kramar1991, Schultze1976, Herrero2001, Lee2010,Sisson2016}  Simultaneously, the theoretical aspects of UPD have been studied from first principles in an effort to connect calculated adsorption energies and model surface structures to experimentally measured underpotential shifts.\cite{Leiva1989,Schmickler1990b,Lehnert1991,Kramar1991,Sanchez1998a,Sanchez1999,Sanchez2001,Greeley2010,Pasti2010,Velez2012}  For a thorough overview of the modeling of UPD phenomena, we direct the reader to the review by Sudha and Sangaranarayanan as well as a more recent review by Oviedo and co-workers.\cite{Sudha2005, Oviedo2015}

The application of first-principles density functional theory (DFT) to model UPD has in several cases been met with great success while simultaneously yielding several perplexing inaccuracies.  One such challenging case to model is the UPD of copper onto the low index surfaces of gold in sulfuric acid media.\cite{Sanchez2001,Greeley2010,Velez2012}
\begin{figure}[H]
	\centering\includegraphics[width=0.92\columnwidth]{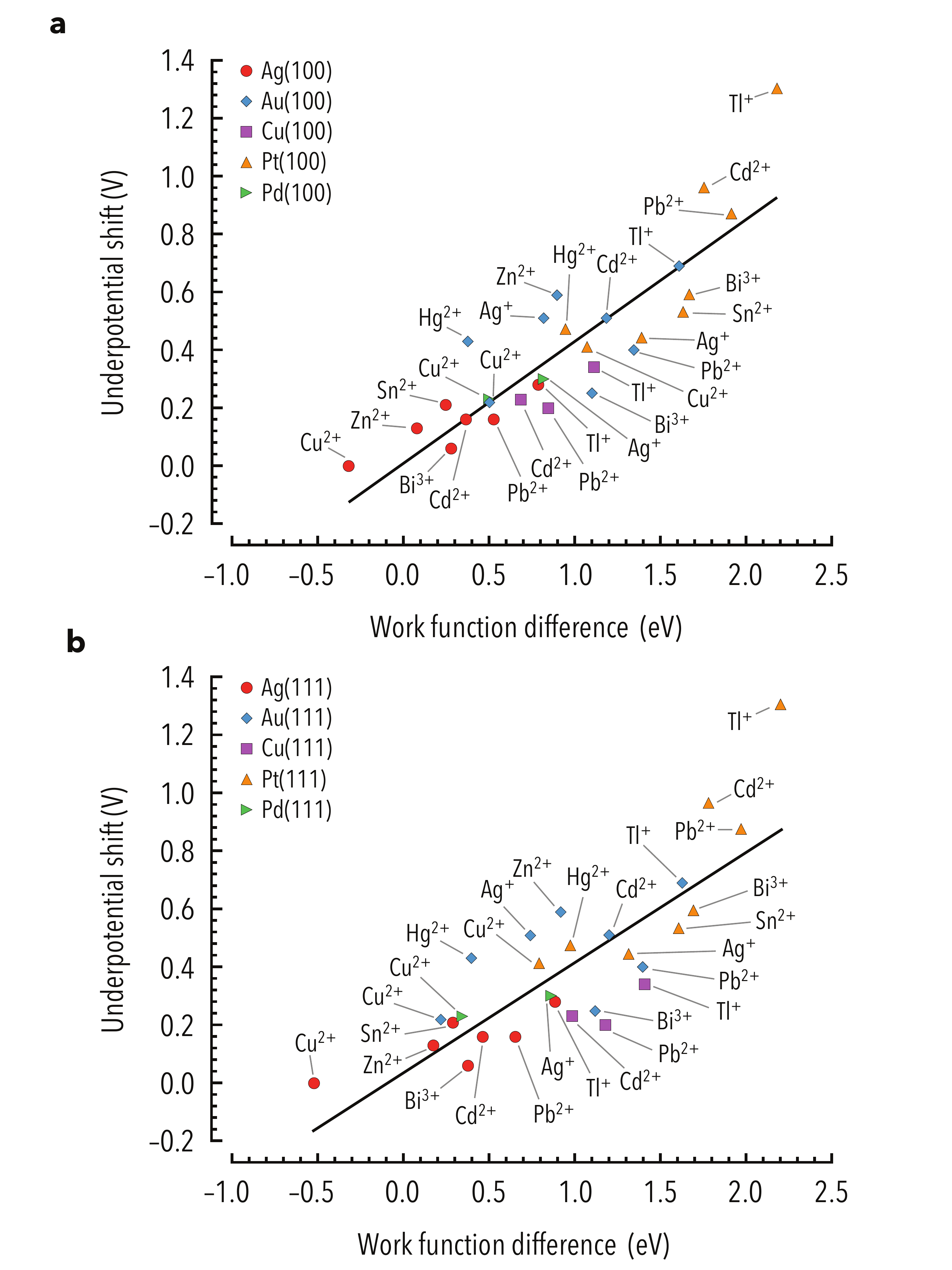}
	\caption{\small Experimental underpotential shifts for polycrystalline electrodes against theoretical single crystal work function differences (see supplementary note 1 for calculation details) for (a) (100) and (b) (111) metal substrates. This correlation underscores the importance of the charge transfer that takes place between the adlayer and the substrate.  }
	\label{fig:interface_PZC}
\end{figure}
While the copper-gold-sulfate system is perhaps one of the most well experimentally characterized UPD couples, first-principles calculations that are performed in vacuum predict copper to desorb at overpotentials, thereby underestimating the voltage-dependent stability of the copper adlayer on gold electrodes.\cite{Sanchez1999,Greeley2010}

Recently, a growing consensus has appeared in the literature that it is necessary to account for the adsorption of  anions to accurately model the stability of metal atoms at the electrode surface.\cite{Gimenez2010, Zhang2014, Velez2012} While these reports represent significant progress in the modeling of interfacial electrochemical phenomena, few investigators have attempted to include other key environmental factors such as the presence of the solvent and the electrical double layer that forms along the interface. Including solvent effects in DFT calculations would deliver more reliable voltage-dependent predictions since the surface charge transfer and the resulting electrostatic energy between the surface dipole and the interfacial electric field would be correctly captured.\cite{Taylor2006, Tripkovic2011, Bjorketun2013}  This energy is, however, a time averaged property of the system that requires one to treat the dynamical motion of the surrounding solvent molecules.  Performing large-scale molecular dynamics simulations of the entire alloy--solution interface is nonetheless computationally demanding due to the algorithmic cost of DFT calculations.  A computationally efficient alternative for calculating surface charge transfer consists of replacing the explicit molecules of the solvent with a polarizable continuum dielectric environment.\cite{Andreussi2012, Fang2013, Letchworth-Weaver2012}  The electrostatic interaction between the surface and the solvent is modeled by computing the response of the system to a polarization density that is induced along the surface of the continuum dielectric cavity surrounding the electrode.  The interfacial dipole potential can thus be obtained without explicitly describing the response of the dynamically evolving solvent, leading to a substantial reduction in computational cost while retaining the essential features of the coupling of the surface dipole with the interfacial electric field.

In this work, we perform a critical analysis of the influence of the electrochemical environment on the voltage-dependent stability of a copper UPD layer on a gold (100) surface. We confirm the prediction of overpotential deposition in vacuum conditions and subsequently demonstrate how experimentally accessible quantities of the electrochemical interface such as the double layer capacitance and the activity of the ions in solution can be taken as environmental parameters in a voltage-sensitive surface stability analysis. We then demonstrate how quantum--continuum results can be used to parameterize grand-canonical electrochemical Monte Carlo simulations that enable us to assess how varying the applied voltage affects the surface composition and surface structure of the UPD adlayer.  Finally, we study the influence of sulfate coadsorption on the UPD of copper on gold (100) under electrochemical conditions.

\section{Results and Discussion}

The voltage dependence of the copper adlayer on a non-reconstructed gold (100) surface is modeled first in a pure continuum solvent and then subsequently in a dilute sulfuric acid medium to account for the presence of sulfate coadsorption. The influence of environmental factors such as the interfacial capacitance and the activity of the hydrated ions is discussed. Details regarding the continuum solvent model are presented below in the methods section.

\subsection{Interfaces under voltage}
 
 Scanning probe measurements indicate that copper forms a pseudomorphic monolayer on the gold (100) surface after the UPD peak in the cathodic scan.\cite{Möller1995, Ikemiya1995} We therefore model the deposition of copper ions onto the four available 4-fold hollow sites of a $2 \times 2$ gold slab surface. The electrodeposition reaction can thus be expressed as 
\begin{equation}
\text{Cu}^{2+}_\text{(aq)} + 2e^- \to \text{Cu}^\ast ,
\end{equation}
for which the accompanying change in free energy is
\begin{equation}
\Delta \mu = \mu_{\text{Cu}^\ast} - (\mu_{\text{Cu}^{2+}} - 2e_0 \Phi),
\label{eq:UPD_free_energy_change}
\end{equation}
where $\mu_{\text{Cu}^\ast}$ is the chemical potential of the adsorbed copper species, $\mu_{\text{Cu}^{2+}} $ is the chemical potential of the hydrated copper ion in solution, and $\Phi$ is the voltage of the gold electrode.  It should be noted that it is challenging to accurately calculate the energy of hydrated ions within the framework of DFT due to the electron delocalization error that plagues conventional exchange-correlation approximations.\cite{Cohen2008}  This problem can be circumvented by employing the definition of the standard reduction potential of copper:

\begin{equation}
\begin{cases}
&\mu_{\text{Cu}^{2+} }   = \mu_{\text{Cu}^{2+} }^\circ + k_\text{B} T \ln a_{\text{Cu}^{2+}} \\
&\mu_{\text{Cu}^{2+} }^\circ   = \mu^\circ_{\text{Cu}}  + 2e_0\Phi^\circ_{\text{Cu/Cu}^{2+}}
\end{cases}
\label{eq:copper_ion}
\end{equation}
where $\mu^\circ_{\text{Cu}}$ is the cohesive energy of bulk copper which we have calculated to be $-3.59$ eV and $\Phi^\circ_{\text{Cu/Cu}^{2+}}$ is the standard reduction potential of copper which has been measured to be 0.34 V vs.~SHE.\cite{Haynes2016} Equation \ref{eq:copper_ion} leads to a standard state copper ion chemical potential of $-2.91$ eV.  Each of these values along with the other thermodynamic data used throughout this work have been summarized below in Table~\ref{tab:thermo_data2}.
\begin{table}[h]
\centering
\caption{ \label{tab:thermo_data2} Thermodynamic data used to define reference energies and chemical potentials. \\}
\begin{ruledtabular}
\begin{tabular}{l r l c l}
Quantity & Value & Unit & Source  &  \\
\hline \\
$\mu^\circ_{\text{Cu}}$                                    & $-$3.59     & eV         & DFT       &  \\
$\mu^\circ_{\text{Cu}^{2+}}$                            & $-$2.91    & eV         & DFT      &  \\
$\mu^\circ_{\text{S}}$                                      &  $-$3.06    & eV        & DFT       & \\
$\mu^\circ_{\text{O}_2}$                                  & $-$6.47     & eV         & DFT       &  \\
$\mu^\circ_{\text{SO}_4^{2-}}$                        & $-$23.67   & eV       & DFT       &  \\
\\
$\Phi^\circ_{\text{Cu/Cu}^{2+}}$                      & 0.34      & V vs.~SHE   & Expt.  & Ref.~\onlinecite{Haynes2016}   \\
$\Phi_\text{Cu, pzc} $                                       & 0.24      & V vs.~SHE   & Expt.   & Ref.~\onlinecite{Hamelin1995}  \\
$\Phi^\circ_{\text{H}_2\text{O} / \text{O}_2}$   & 1.23     & V vs.~SHE   & Expt.    & Ref.~\onlinecite{Haynes2016}   \\
$\Phi^\circ_{\text{S/HSO}_4^-}$                      & 0.34       &V vs.~SHE     & Expt.    & Ref.~\onlinecite{Bouroushian2010} \\
\\
p$K_\text{a}$(HSO$_4^-$)                                                   & 1.99       &                & Expt.    & Ref.~\onlinecite{Haynes2016} \\ 
\\
$C_\text{dl}(\theta=0.00)$                                & 14.43        & $\mu$F/cm$^2$     & DFT       &  \\
$C_\text{dl}(\theta=0.25)$                                & 20.73        & $\mu$F/cm$^2$     & DFT        & \\
$C_\text{dl}(\theta=0.50)$,  $p(2\times1)$          & 20.39        & $\mu$F/cm$^2$     & DFT     & \\
$C_\text{dl}(\theta=0.50)$,  $c(2\times2)$        & 21.16       & $\mu$F/cm$^2$     & DFT        & \\
$C_\text{dl}(\theta=0.75)$                                & 20.21        & $\mu$F/cm$^2$     & DFT        &  \\
$C_\text{dl}(\theta=1.00)$                                & 19.26        & $\mu$F/cm$^2$     & DFT        & 
\end{tabular}
\end{ruledtabular}
\end{table}
Compared to the ion in solution, the chemical potential of the adsorbed copper species requires a few additional considerations, namely the number of copper atoms present on the surface unit cell and the charge on the electrode surface.   We determine the chemical potential of the copper adatom by first computing the free energy of the neutral copper covered surface with respect to the number of copper adsorbates $N$, giving us the binding energy $F_0(N)$. As indicated by the data in Table~\ref{tab:thermo_data}, the adsorption energy exhibits a near-linear dependence on the surface coverage. Expanding this energy with respect to the total charge $Q$, we obtain the free energy of the surface unit cell as a function of the adsorbate number and the surface charge
\begin{equation}
F(N, Q) = F_0(N) + \Phi_0 (N)Q + \frac{1}{2} \frac{Q^2}{C_\text{dl}},
\label{eq:free_energy}
\end{equation}
where $\Phi_0 (N)$ is the potential of zero charge of the surface with $N$ copper atoms adsorbed (Table~\ref{tab:thermo_data}), and $C_\text{dl}$ is the double layer capacitance of the interface.  The potential of zero charge is calculated following the approach detailed in the methods section, where we take the converged electrostatic potential near the edge of the cell as the reference, as shown in Fig.~2.
\begin{figure}[t]
	\centering\includegraphics[width=0.95\columnwidth]{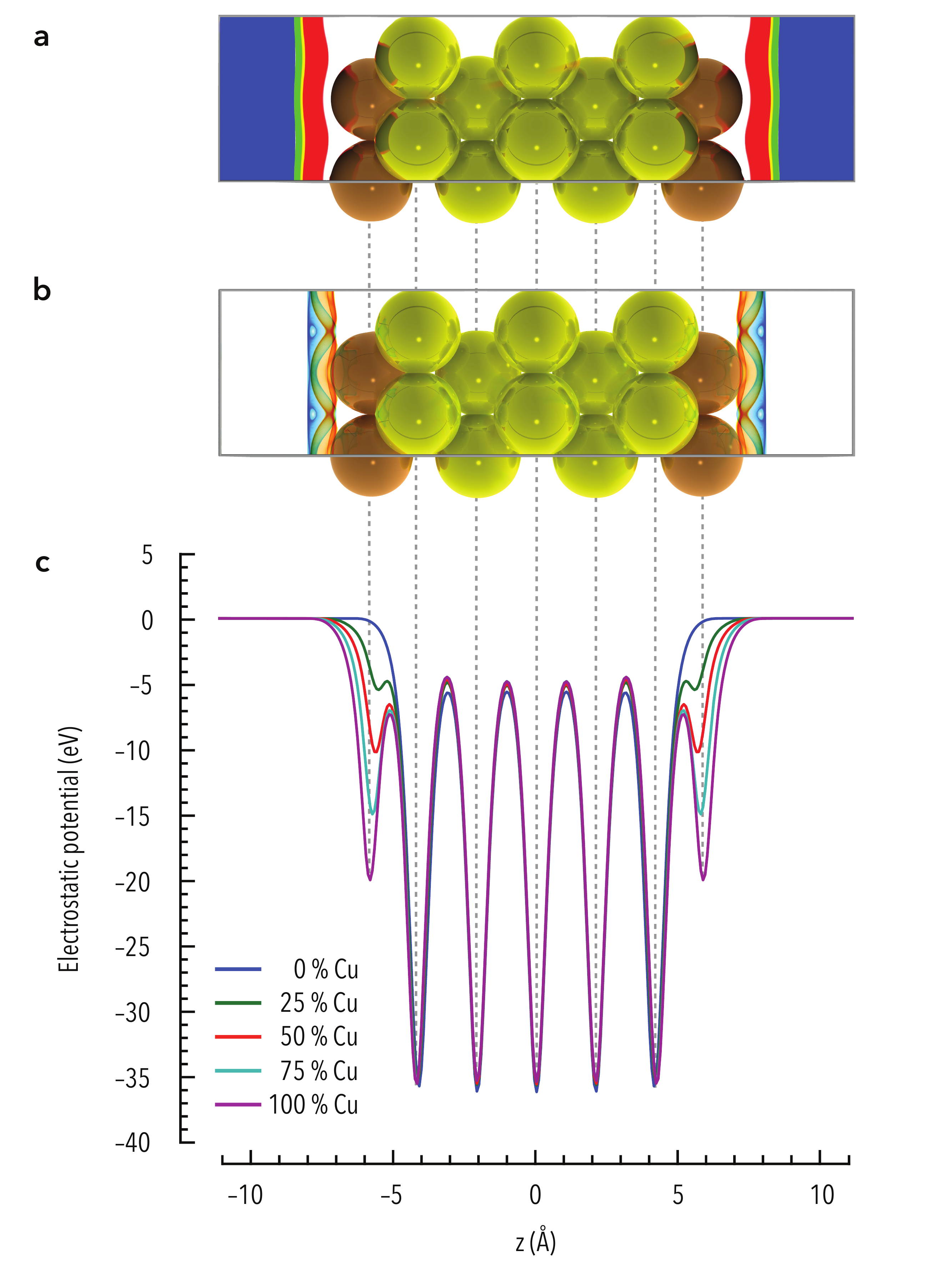}
	\caption{\small (a) Dielectric cavity of the continuum solvent. The transparent region is associated with $\epsilon_0=1$ which transitions outside of the electrode surface to the blue region where $\epsilon_0 = 78.3$. (b) Polarization charge density that arises along the surface of the dielectric cavity. Positive polarization charges are shown in red, negative charges in blue. (c) The electrostatic potential for each copper covered surface is aligned to zero in the bulk of the continuum solvent region. The perturbation of the outermost peak is related to the change in surface dipole moment due to the adsorption of copper ions for increasing coverages.}
	\label{fig:voltage_extraction}
\end{figure}
\begin{table}[b]
\centering
\caption{ \label{tab:thermo_data} Thermodynamic data obtained from the quantum--continuum calculations as a function of the number of adsorbed copper species. See supplementary note 3 for details. \\}
\begin{ruledtabular}
\begin{tabular}{c c c c c}
$N$ & Structure & $F_0$ (eV) & $\Phi_0$ (V)  & $\Phi_0$ (V vs.~SHE) \\
\hline \\
0 & $p(1\times1)$ & $\ \ $0.00 & 4.88 & $\ \ $0.24 \\
1 & $p(2\times2)$ & $-$3.29 & 4.37 & $-$0.27 \\
2 & $p(2\times1)$  & $-$6.75 & 4.27 & $-$0.37 \\
2 & $c(2\times2)$  & $-$6.43 & 3.93 & $-$0.71\\
3 & $p(2\times2)$ & $-$10.05 & 4.05 & $-$0.59  \\
4 & $p(1\times1)$ & $-$13.51 & 4.06 & $-$0.58
\end{tabular}
\end{ruledtabular}
\end{table}

Here, we note that the capacitance can be calculated from first principles by computing the energy of the electrode surface with a range of surface charges and subsequently fitting Eq.~\ref{eq:free_energy}. In this approach, a planar ionic countercharge is introduced in the continuum solvent region 3--5 \AA\ from the surface to compensate the surface charge. 
As an example, we have computed the interfacial capacitance at each coverage for a countercharge distance of 3 \AA\ and report the results in Table~\ref{tab:thermo_data2}. We find that the capacitance is relatively insensitive to the degree of copper coverage, with the largest difference being for the pure gold (100) surface and the gold surface with a coverage of $\theta = 0.25$, where the double layer capacitance increases from 14.43 $\mu$F/cm$^2$ to 20.73 $\mu$F/cm$^2$. Both of these values are in close agreement with the experimentally determined double layer capacitance near the potential of zero charge for the gold (100) surface in dilute sulfuric acid media.\cite{Hamelin1995}  It should be mentioned that when using a planar ionic countercharge (the Helmholtz model), the capacitance is a constant with respect to the applied voltage. Other models, such as the Gouy--Chapman model, employ a diffuse ionic countercharge that leads to a voltage-dependent double layer capacitance. While these different models provide a detailed description of the electrochemical interface, we note that the overall goal of this study is to perform a sensitivity analysis of the interfacial capacitance on underpotential deposition phenomena. As a result, the capacitance will serve as an environmental parameter that controls the possible range of copper deposition for experimentally relevant electrolytic conditions. We have thus studied the influence of capacitance values ranging from 0 -- 100 $\mu$F/cm$^2$ for our analysis.  

After specifying the capacitance, the chemical potential can be computed by taking the derivative of Eq.~\ref{eq:free_energy} with respect to $N$, yielding
\begin{equation}
\begin{split}
\mu_{\text{Cu}^\ast} (N, \Phi) &= F_0(N) - F_0(N -1) \\
                                               & \quad + \big( \Phi_0 (N) -  \Phi_0 (N-1)\big)Q_\Phi ,
\end{split}
\end{equation}
where the charge on the surface is calculated with respect to the average of the potentials of zero charge for the surface with $N$ and $N -1$ adsorbed coppers
\begin{equation}
Q_\Phi = C_\text{dl} \left( \Phi - \frac{1}{2} \big( \Phi_0 (N) + \Phi_0 (N -1) \big) \right) .
\end{equation}

With these results in hand, we can now parameterize a voltage-dependent Ising model of the copper--covered gold surface whose energy reads
\begin{equation}
\label{eq:linear_monte_carlo}
\begin{split}
  F(\{\sigma_i\},\Phi) &=\frac 12 \Delta\mu \Big(\theta= \frac 12, \Phi \Big) \sum_i \sigma_i  \\
  &\quad + \frac 1{4z} \frac{\partial \Delta \mu}{\partial \theta}\Big(\theta=\frac 12, \Phi \Big)  \sum_i  {\sum_j}^\prime \sigma_i \sigma_j,
  \end{split}
\end{equation}
where the double summation is restricted to the $z$ nearest neighbors of each lattice site.  In this model, the variable $\sigma_i$ describes the state of the surface; it equals $+ 1$ when the $i$th lattice site is occupied by copper, whereas it equals $-1$ when the lattice site is vacant.  The onsite energy and interaction energy parameters are determined from a linear interpolation of the chemical potential between $\Delta \mu(N = 1, \Phi)$ and $\Delta \mu(N = 4, \Phi)$:
\begin{equation}
\begin{split}
\Delta\mu(\theta&, \Phi) = \frac{ \Delta \mu(N = 1,\Phi) + \Delta \mu(N = 4,\Phi) }{2} \\
                                    &+  \left(2\theta -1 \right) \frac{ \Delta \mu(N = 4,\Phi) -  \Delta \mu(N= 1,\Phi)  }{2},
\end{split}
\label{eq:interp_chem_pot}
\end{equation}
which describes the coverage-dependent cost to transfer a copper ion from solution onto the gold surface.  The acceptance or rejection of a trial adlayer is governed by the Metropolis algorithm, for which the acceptance rate is set by the temperature of the system.  Each simulation was performed at a constant temperature of 298 K and at a set of fixed voltages to enable the construction of adsorption isotherms.

Finally, it should be mentioned that the correspondence between the absolute potential scale of the Monte Carlo calculations and the standard hydrogen electrode reference of electrochemical measurements can be determined unambiguously by comparing the predicted potential of zero charge of the substrate in the continuum solvent with the experimental potential of zero charge for the same electrode in solution (Table~\ref{tab:thermo_data}). In this case, for  the gold (100) surface, the potential of zero charge has been measured to be approximately 0.24 V vs.~SHE in dilute sulfuric acid media.\cite{Hamelin1995} The theoretical scale is then shifted by this difference in such a way that the charge of the surface at any given voltage in the vicinity of the potential of zero charge is correctly described.

\subsection{Effect of the applied voltage}

 \begin{figure*}[t]
	\centering\includegraphics[width=0.95\linewidth]{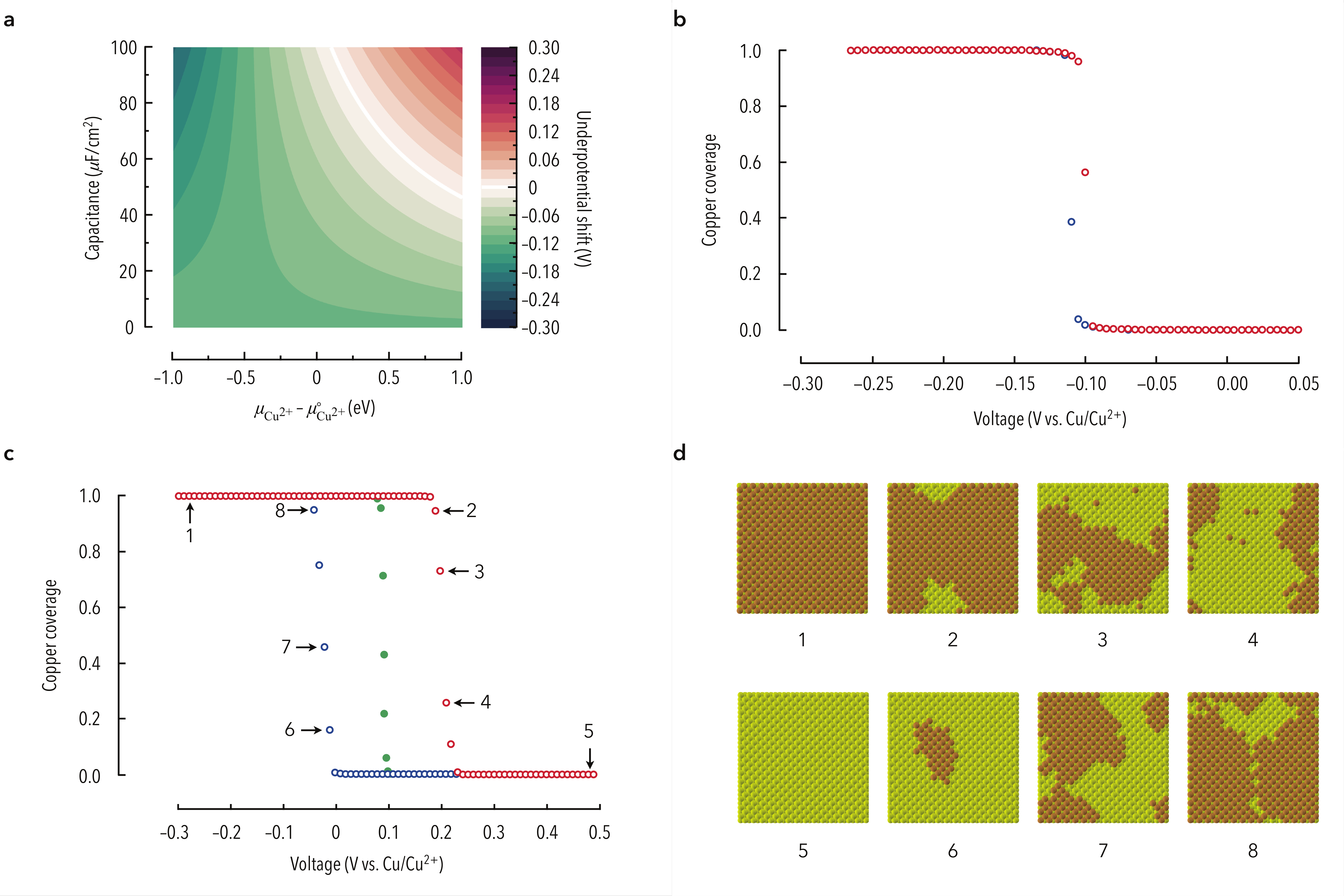}
	\caption{\small Electrochemical Monte Carlo results for copper UPD on the gold (100) surface. (a) The computed underpotential shift for a half-covered gold (100) surface becomes more positive with increasing double layer capacitance and copper ion chemical potential. (b) The adsorption isotherm of the gold surface under vacuum conditions indicates the copper adlayer is stable up to $-0.11$ V vs. Cu/Cu$^{2+}$. (c) The adsorption isotherm of the gold surface with a capacitance of 80 $\mu$F/cm$^2$ and a copper ion chemical potential 1 eV higher than the standard state. The red curve (labels 1$\to$5) is for the run with an initially complete monolayer, and the blue curve (labels 5$\to$8$\to$1) is for the run with an initially pristine gold surface. The green curve shows the physical isotherm, when all surfaces are initialized with a random coverage of $\theta = 0.5$.  (d) Simulation snapshots along the cathodic and anodic voltage scans associated with the isotherm in panel c. }
\label{fig:Au_Cu_MC_results}
\end{figure*}

In order to elucidate the voltage-dependent stability of the adsorbed coppers at the gold (100) surface, we have performed two sets of Monte Carlo simulations: one for which the capacitance is set to 0 $\mu$F/cm$^2$ and the copper ions are under standard conditions, and another for which the capacitance is set to 80 $\mu$F/cm$^2$ and the chemical potential of the copper ions has been shifted by 1 eV.  The first set of environmental parameters were chosen in order to verify that the Monte Carlo model reproduces the literature theoretical underpotential shift value of $-0.11$ V in vacuum (that is, no capacitive effect) conditions.\cite{Sanchez1999} The second set of parameters were chosen based on a stability map that was produced by analytically solving Eq.~\ref{eq:interp_chem_pot} for the underpotential shift when the adsorbed coppers on a half-covered gold surface were in equilibrium with the solution, as shown in Fig.~3a (see supplementary note 2 for more details).  The stability map indicates that there is a clear transition between positive and negative underpotential shifts, and that by considering the capacitance of the interface, the stability of the adsorbed coppers at the surface can be enhanced, pushing the computed underpotential shift to more positive values. Similarly, larger activities for the hydrated copper ion also enhances the stability of the adsorbed copper species.

The Monte Carlo simulations were performed by initially running two sets of calculations, one starting with a full copper monolayer and the second with a pristine gold surface to simulate the initial state of the surface during an anodic and a cathodic voltage scan, respectively. We averaged the voltage--coverage results across 100, 200, 500, 1000, and 5000 runs in a sensitivity analysis in an effort to minimize the uncertainty in the computed isotherms. We found that there was a negligible difference between the results obtained from averaging over 100 runs and 5000 runs, so we opted for the smaller number of runs for the sake of computational expediency. We also note that the coverage distribution computed at each voltage was exceedingly narrow far from the phase transition with a standard deviation lower than $10^{-4}$; however, the spread of the coverage results broadened when the applied voltage was within a range of $\pm 0.03$ V of the phase transition with a maximum standard deviation of 0.45 at the phase transition voltage. This trend was consistent across all of the tests that were ran, independent of the number of runs the results were averaged over, which further supported our decision to average future simulations over 100 runs. 

When the Monte Carlo simulations are performed under vacuum conditions (Fig.~3b), we obtain a sharp and complete depletion of the surface at $-0.11$ V relative to the standard redox potential of copper.  The sharp transition is indicative of a phase transition on the surface at a voltage that is consistent with the previously computed value in the literature.  In contrast, when the capacitance of the interface and the chemical potential of the hydrated ion are considered (Fig.~3c) the electrochemical Monte Carlo simulations exhibit a shift in the phase transitions to potentials more positive than the copper redox potential, thereby yielding a more qualitatively accurate description of the voltage-dependent stability of the adsorbed copper atoms.  We also note that a hysteresis in the voltage--coverage isotherm develops, indicating that the copper adatoms are more stable when they are fully coordinated within the full surface monolayer than they are for an initially bare surface. In principle, this type of thermodynamic hysteresis that depends upon the initial coverage occurs when two locally stable phases are separated by an unstable region along the isotherm. The phase transition thus occurs at a fixed voltage where multiple stable states have become available to the system, and the system spontaneously moves to the state with the lower free energy. It should be noted that the physical phase transition occurs at the voltage where the free energy vs. chemical potential curve intersects itself, and at any other voltage, the phase transitions appear as a consequence of metastability. The physical voltage--coverage isotherm, as shown in Fig.~3c with the filled markers, can be recovered by initializing the surface simulations with a randomly dispersed half-monolayer of copper on the gold (100) surface, placing the system into an initially unstable state. The latter creates a driving force for the system to attain a more stable coverage; however in this case, the system will not encounter another unstable region along its trajectory unlike the systems initialized with zero or full coverage.  The result is a single phase transition that occurs between the two hysteretic phase transitions connecting stable states with equal chemical potentials. This is equivalent to the result one would find by constructing the common tangent plane connecting the two stable states of the free energy surface. Nevertheless, it is important to investigate the metastable states of the system since they correspond to the physical conditions of the gold surface during a voltammetric experiment.  This analysis revealed that the copper adatoms appear to be stable as a pseudomorphic monolayer on the gold (100) surface (Fig.~3d), consistent with the copper layers that form in electrolytes with low bisulfate concentrations.\cite{Möller1995, Ikemiya1995}

In view of these results, it is clear that accounting for the capacitive nature of the interface and the composition of the solution can enhance the description of the voltage-dependent stability of the electrode surface. This model does, however, take a somewhat simplistic view of the UPD reaction, since it has been suggested in both the experimental and theoretical literature that anion coadsorption can play a significant role in stabilizing the UPD layer. For example, it has recently been shown for copper UPD on a gold (111) surface that the inclusion of coadsorbed sulfate ions can yield more reliable underpotential shift predictions in vacuum.\cite{Velez2012} It has also been speculated that sulfate could adsorb alongside copper on the gold (100) surface, however, this has yet to be examined in the theoretical literature.\cite{Möller1995, Ikemiya1995} Therefore, in the following section, we carry out one of the first analyses of the effects of sulfate coadsorption on the UPD of copper on the gold (100) surface under electrochemical conditions.

\subsection{Effect of anion coadsorption}
In order to account for the presence of sulfate in the system, we consider the formation of a $\emph{p}(2\times 2)$ overlayer of sulfates occupying the 4-fold hollow sites of the copper adlayer as shown in Fig.~4a. We must therefore consider the new electrodeposition process
\begin{equation}
4\text{Cu}^{2+}_\text{(aq)} + \text{SO}^{2-}_{4\text{(aq)}} + 6e^- \to \text{Cu}_4(\text{SO}_4)^\ast.
\end{equation}
The new equilibrium condition is specified by the following electrochemical potential balance
\begin{equation}
4\mu_{\text{Cu}^{2+}} + \mu_{\text{SO}_4^{2+}}  - 6e_0 \Phi_\text{UPD} = \mu_{\text{Cu}_4(\text{SO}_4)^\ast}(  \Phi_\text{upd} ) ,
\label{eq:UPD_equilibrium}
\end{equation}
where we employ the method presented in the previous section to derive the chemical potential of sulfate from fundamental thermodynamic identities. To this end, we consider the following equilibrium relations:
\begin{equation*}
\begin{cases}
&\mu^\circ_{\text{H}^+} -e_0 \Phi^\circ_{\text{H}_2 / \text{H}^+} = \frac{1}{2} \mu^\circ_{\text{H}_{2}}  \\
&\mu^\circ_{\text{H}_2\text{O}} = \mu^\circ_{\text{H}_2} + \frac{1}{2} \mu^\circ_{\text{O}_2} - 2e_0 (\Phi^\circ_{\text{H}_2\text{O} / \text{O}_2}  - \Phi^\circ_{\text{H}_2 / \text{H}^+}) \\
&\mu^\circ_{\text{HSO}_4^-} = \mu^\circ_{\text{H}^+} + \mu^\circ_{\text{SO}_4^{2-}} - k_\text{B}T\ln(10) \text{p}K_a   \\
&\mu^\circ_\text{S} + 4\mu^\circ_{\text{H}_2\text{O}} = \mu^\circ_{\text{HSO}_4^-} +7\mu^\circ_{\text{H}^+} - 6e_0 \Phi^\circ_{\text{S/HSO}_4^-} ,
\end{cases}
\end{equation*}
which allows us to write
\begin{equation}
\begin{split}
\mu^\circ_{\text{SO}_4^{2-}} &= \mu^\circ_\text{S} + 2 \mu^\circ_{\text{O}_2} - 8e_0\Phi^\circ_{\text{H}_2\text{O} / \text{O}_2} \\
       &\quad + 6e_0\Phi^\circ_{\text{S/HSO}_4^-} + k_\text{B}T\ln(10) \text{p}K_a
\end{split}
\end{equation}
where $\mu^\circ_\text{S}$ is the cohesive energy of bulk sulfur which we have computed to be $-3.06$ eV for the crystalline $\alpha$-phase,  $\mu^\circ_{\text{O}_2}$ is the free energy of an isolated oxygen molecule which is computed to be $-6.47$ eV, $\Phi^\circ_{\text{H}_2\text{O} / \text{O}_2}$ and $\Phi^\circ_{\text{S/HSO}_4^-}$  are the standard reduction potentials of water and sulfur which have been measured to be 1.23 and 0.34 V vs.~SHE, respectively, and finally p$K_a$ is the negative log of the bisulfate acid dissociation constant which has been measured to be 1.99.\cite{Bouroushian2010,Haynes2016}  The free energy of the sulfate anion under standard conditions is thus computed to be $-23.67$ eV. 

As in the case of the previous section, the underpotential shift can be computed analytically with respect to the interfacial capacitance and the chemical potentials of the ions in solution. 
\begin{figure}[h!]
	\centering\includegraphics[width=0.95\columnwidth]{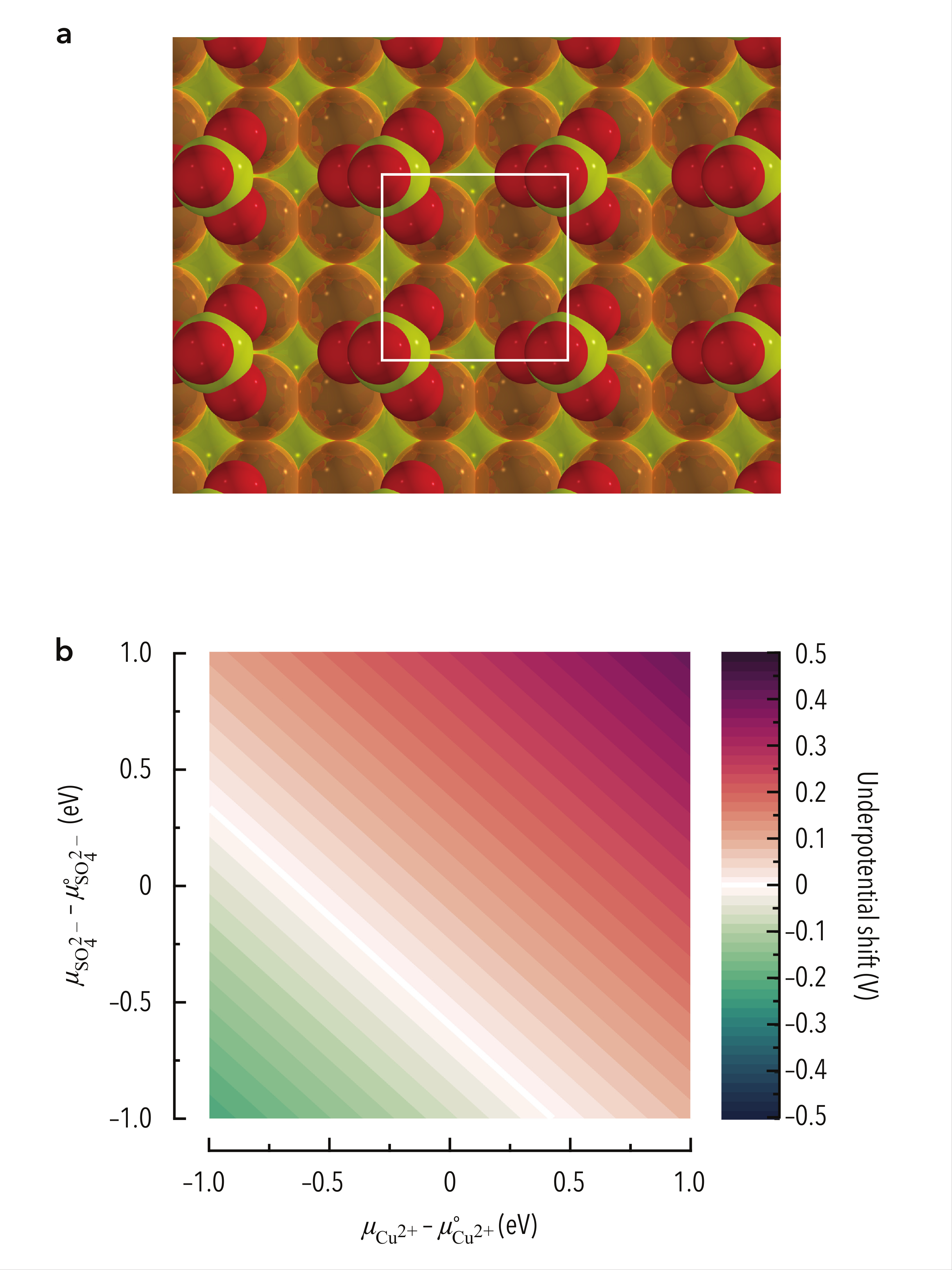}
	\caption{\small (a) The co-adsorbed sulfates are included as a \emph{p}(2$\times$2) overlayer at the 4-fold hollow sites of the copper adlayer. (b) Influence of the activities of the hydrated sulfate ion and hydrated copper ion on the underpotential shift at a capacitance of 20 $\mu$F/cm$^2$. 
  }
	\label{fig:sulfate_co_adsorption}
\end{figure}
When the capacitance is set to 0 $\mu$F/cm$^2$ and the ion activities are set to unity, the underpotential shift attains a positive value of 0.07 V, indicating that the presence of the co-adsorbed sulfates plays an important role in stabilizing the copper adlayer on the gold (100) surface.  This stability is again enhanced, however, when the capacitance of the interface is set to 20 $\mu$F/cm$^2$ (Fig.~4b) which is approximately the capacitance at the potential of zero charge of the gold (100) surface in dilute sulfuric acid solutions,\cite{Hamelin1995} leading to an underpotential shift of 0.10 V.  The underpotential shift increases with increasing capacitance values to 0.12 V and  0.14 V for capacitances of 40 and 60 $\mu$F/cm$^2$, respectively, all within the experimentally observed underpotential shift range.\cite{Möller1995} Clearly, the capacitance of the interface can play an important role in stabilizing the UPD layer. In the same vein, concentration effects also play a critical role in stabilizing the adsorbed species.  As shown in Fig.~4b, when the capacitance is considered to be fixed at 20 $\mu$F/cm$^2$, if the chemical potential of copper increases by 0.25 eV, the UPD shift increases to 0.14 V from 0.10 V, and if the chemical potential of sulfate is simultaneously increased by 0.25 eV, the UPD shift attains a value of 0.18 V.  Therefore, it appears that anion coadsorption, capacitive effects, and concentration effects can operate cooperatively to give rise to the observed stability of the copper UPD layer on the gold (100) surface.

\section{Summary}
We have studied computationally the influence of the applied voltage and the composition of the electrolyte on the stability of an underpotentially deposited copper adlayer on a gold (100) surface. We have explored the use of quantum--continuum calculations via the self--consistent continuum solvation model to parameterize a two-dimensional Ising model for electrochemical Monte Carlo simulations of UPD reactions.\cite{Andreussi2012} By incorporating experimentally accessible environmental parameters such as the double-layer capacitance of the electrode--electrolyte interface and the chemical potentials of the ions in solution, a computationally efficient voltage--sensitive stability analysis has been demonstrated to predict both the equilibrium structures of adsorbate-covered surfaces and the coverage-dependence of the equilibrium voltage on the electrode under electrochemical conditions.

We have furthermore confirmed that gas-phase DFT calculations predict an overpotential shift for copper electrodeposition on the (100) surface of gold. With the aid of the comprehensive interfacial model proposed in this work, it has been shown that more reliable predictions of the underpotential shift can be obtained by considering the influence of specific environmental features of the electrochemical interface such as the presence of the solvent, the interfacial capacitance, anion coadsorption, as well as the composition and concentration of the electrolyte. We believe the model developed herein will be useful in atomistic first-principles studies of UPD reactions and further enable the detailed study of a broad class of interfacial electrochemical processes.

\section{Computational methods}

\label{sec:computational_methods}

Quantum--continuum calculations are carried out using the {\sc pwscf} code of the open-source {\sc quantum-espresso} software with the newly released {\sc environ} module.\cite{Giannozzi2009, Andreussi2012}  We adopt a slab model to represent the electrode surfaces, whereby the adlayer is included symmetrically at the top and bottom layers to minimize spurious dipole interactions across the supercell.  We consider slabs of seven layers in the $2 \times 2$ primitive cell geometry.  We employ the Perdew--Burke--Ernzerhof (PBE) exchange-correlation functional to describe quantum electronic interactions with ultrasoft pseudopotentials to represent the ionic cores.  We set the kinetic energy and charge density cutoffs to be of 40 Ry and of 480 Ry, respectively, after verifying numerical convergence of the interatomic forces within a few meV/\AA\ and of the total energies within 50 meV per cell. The Brillouin zone is sampled with a shifted $4\times 4 \times1$ Monkhorst--Pack grid, and the electronic occupations are smoothed with 0.02 Ry of Marzari--Vanderbilt cold smearing.  The slabs are centered in each cell and it is found that a vacuum height of 10 {\AA} was sufficient to converge the electrostatic potential at the cell boundaries with the generalized electrostatic solvers that have been implemented in the module.\cite{Dabo2008, Andreussi2014}

Solvent effects are included by computing the response of the continuum dielectric medium along a smooth dielectric cavity that is constructed on the self-consistently calculated electron density of the surface (the self-consistent continuum solvation model). \cite{Andreussi2012} This approach is inspired by the solvation model of Fattebert and Gygi with the main difference that the density-dependent parameterization of the dielectric cavity is logarithmically smooth instead of being linearly smooth, thereby enabling the convergence of surface simulations where the electron density can exhibit sharp fluctuations at the interface.\cite{Fattebert2002, Fattebert2003} In specific terms, the shape of the dielectric cavity is controlled by an inner and an outer isocontour surface of the electron charge density, $\rho_\text{max}$ and $\rho_\text{min}$, respectively.  The density dependence of the dielectric permittivity is thus defined as $\epsilon(\rho) = \exp[( \zeta_\rho - \sin(2\pi \zeta_\rho) / 2\pi)\ln \epsilon_0]$, where $\epsilon_0$ is the dielectric permittivity of the surrounding solvent and the density-dependent $\zeta_\rho$ variable is defined as $\zeta_\rho = (\ln \rho_\text{max} - \ln \rho)/(\ln \rho_\text{max} - \ln \rho_\text{min})$. The self-consistent continuum solvation parameterization also incorporates non-electrostatic cavitation contributions such as the external pressure, solvent surface tension, as well as solvent dispersion and repulsion effects,\cite{Andreussi2012} where the latter two are expressed as $G_\text{cav} = \gamma S$ and $G_\text{dis+rep} = \alpha S + \beta V$. Here, $\gamma$ is the experimental solvent surface tension while $\alpha$ and $\beta$ are fitted parameters. $S$ and $V$ are the quantum surface and quantum volume of the solute, respectively, and are defined as
$S = -\int d\bm{r} \frac{ d\Theta}{d\rho} (\rho) \vert \nabla \rho \vert$
and $V = \int d\bm{r} \Theta(\rho)$
for which the density-dependent $\Theta$ function is written $\Theta(\rho) = (\epsilon_0 - \epsilon(\rho) ) / ( \epsilon_0 - 1)$.\cite{Andreussi2012, Cococcioni2005}
Together, the electrostatic and non-electrostatic contributions to the solvation free energy require two solvent-dependent parameters which can be taken from experiment, namely $\epsilon_0$ and $\gamma$, and four tunable parameters which can be obtained by fitting against a set of known experimental solvation energies: $\rho_\text{min}$, $\rho_\text{max}$, $\alpha$, and $\beta$. Recently, parameterizations have been published for neutral and charged molecules in aqueous solvents which can reproduce experimental solvation energies with mean absolute errors of 1.3 kcal/mol for neutral molecules, 2.27 kcal/mol for cations, and 5.54 kcal/mol for anions.\cite{Andreussi2012, Dupont2013}

Employing the parameterization of Andreussi \emph{et al.} for neutral species ($\epsilon_0 = 78.3$, $\rho_\text{max} = 5\times 10^{-3}$ a.u., $\rho_\text{min} = 1\times 10^{-4}$ a.u., $\gamma = 72.0$ dyn/cm, $\alpha=-22$ dyn/cm, $\beta = -0.35$ GPa),\cite{Andreussi2012} this model enables us to compute the free energy of the surface by relaxing the slab under the constant pressure applied by the continuum environment and for constant surface charge.  By varying the number and configuration of copper adsorbates on the gold (100) surface in the simulation cell, the dependence of the free energy on the surface coverage and the UPD layer structure can be determined. Finally, after obtaining the relaxed electrode surface structure, we compute the voltage in each calculation by taking the converged electrostatic potential in the bulk of the continuum solvent, far away from the interface, as the electrostatic reference. By setting this reference, the voltage can be directly computed as the opposite of the Fermi level of the electrode.

\begin{acknowledgements}
This work was funded by the Soltis Faculty Support Gift, the Ralph E. Powe Junior Faculty Award from Oak Ridge Associated Universities, and by the Center for Dielectrics and Piezoelectrics at Penn State University. The authors thank the Penn State Institute for CyberScience for providing high-performance computing resources and technical support throughout this work.
\end{acknowledgements}

\section*{Competing Interests}
The authors declare no conflict of interest.
\vspace{24pt}
\section*{Contributions}
SW and ID contributed significantly to all aspects of the work.

\end{document}